\documentstyle[prc,preprint,tighten,aps]{revtex}
\begin {document}
\draft
\title{
Microscopic description of the beta delayed deuteron emission
from \bbox{^6}He}
\author{Attila Cs\'ot\'o$^{1,2}$\cite{email} and Daniel Baye$^1$}
\address{
$^1$Physique Nucl\'eaire Th\'eorique et Physique Math\'ematique,
C.P. 229, Campus Plaine, \\
Universit\'e Libre de Bruxelles, B--1050 Brussels, Belgium\\
$^2$Institute of Nuclear Research of the Hungarian Academy of
Sciences \\ P.O.Box 51 Debrecen, H--4001, Hungary}
\date{\today}

\maketitle

\begin{abstract}
\noindent
The beta delayed deuteron emission from $^6$He is studied in a
dynamical microscopic cluster model. This model gives a
reasonably good description for all the subsystems
of $^6$He and $^6$Li in a
coherent way, without any free parameter.
The beta decay transition probability to the $^6$Li ground state is
underestimated by a few percents.
The theoretical beta delayed deuteron spectrum is close to
experiment but it is also underestimated by about a factor 1.7.
We argue that, in spite of their different magnitudes, both
underestimations might have a common origin. The model confirms
that the neutron halo part of the $^6$He wave function plays a
crucial role in quenching the beta decay toward the $\alpha$ + d
channel.
\end{abstract}
\pacs{PACS numbers: 23.40.Hc, 21.60.Gx, 27.20.+n}

\narrowtext

\section{Introduction}

In the past few years the large amount of experimental data which
have been accumulated on the structure and reactions of unstable
nuclei revealed the existence of a neutron halo structure in
nuclei with a large neutron excess \cite{Tanihata,Hansen}. The
best known of these nuclei is $^{11}$Li. However, there are
ambiguities in the theoretical description of this nucleus
mainly because few $^9$Li + n scattering data are available, and
because the $^9$Li core is soft \cite{DanilinPR}. Fortunately,
these problems do not occur in the case of the other prominent
neutron halo nucleus, $^6$He. Sophisticated
calculations exist which can account for several properties of
$^6$He \cite{DanilinPR,Kukulin,Lehman,Suzuki,Csoto}.

One of the most surprising feature of $^6$He is the very small
Gamow-Teller (GT) beta decay branching ratio toward the $\alpha$ + d
continuum channel. The $\beta$ delayed deuteron emission from $^6$He
was first observed in Ref.\ \cite{Riisager}, and the measured
branching ratio $(2.5\pm 0.5)\cdot 10^{-6}$ is two orders of magnitude
smaller than the result of a phenomenological analysis of
the decay process in terms of an $R$-matrix formalism \cite{Riisager}.
Later an experiment with higher statistics resulted in a new value
$(7.6\pm 0.6)\cdot 10^{-6}$ \cite{Borge}.
Even this larger value is overestimated
in a potential model \cite{Descouvemont} and in a
three-body $\alpha$ + n + n model \cite{Zhukov}.

A semimicroscopic model \cite{Baye} shows that the smallness of
the branching ratio is the result of a cancellation,
taking place between two parts of the GT matrix
element, which have different signs. The 'internal' part comes
from the typical nuclear regions ($r < 5$ fm) of the $^6$He
and $\alpha$ + d wave functions. The 'external' part comes from the
5 fm $< r <$ 20 fm region. The importance of the latter part
is a clear consequence of the halo structure in $^6$He.
In that semimicroscopic model, the $^6$He initial state is described
with a 6-nucleon microscopic wave function but the final state is
treated in a potential model. This hybrid treatment is flexible enough
to reveal the quenching mechanism of the branching
ratio. However, its agreement with experiment is probably partly
fortuitous because the results are very sensitive to model assumptions.
In Ref. \cite{Descouvemont} the branching ratio is overestimated
by an order of magnitude because the assumption of a pure $\alpha$
+ dineutron configuration enhances the 'external' part of the GT matrix
element. In Ref. \cite{Zhukov} the overestimation is reduced because
of a much more realistic description of $^6$He but the
treatment of the $\alpha$ + d scattering state is not consistent
with the treatment of the ground state of $^6$He.

Clearly, an accurate and consistent description of this process is
desirable. However, obtaining reliable
results requires that the following conditions are met. The model
must reproduce in a realistic way (i) the deuteron
binding energy and size, (ii) the low-energy n + n phase shifts, (iii) the
$\alpha$ + n scattering, (iv) the $\alpha$ + d scattering, and (v)
the $^6$He binding energy and size. In addition the analysis
of Ref. \cite{Baye} indicates that (vi) the halo must be well
described up to distances as large as 20 fm.
In this paper we calculate the $\beta$ delayed deuteron spectrum in
a fully microscopic model. We try to use a parameter free model
without any {\em ad hoc} model assumption. With the Minnesota interaction
conditions (i) to (iv) are fulfilled \cite{Minnesota,Fujiwara} and
a good $^6$He wave function satisfying (v) is available \cite{Csoto}
although the basis has to be enlarged to meet (vi). The fact
that a few percent error in the model at typical nuclear scales
can modify the order of magnitude of the final result, as
is pointed out in Ref. \cite{Baye}, calls for a test example which is
sensitive to the details of the wave functions only up to less than
10 fm. Our test case will be the $\beta$ decay transition
to the ground state of $^6$Li.

The model is recalled in Sec. II. The results are described
and discussed in Sec. III. Conclusions are presented in Sec. IV.
\section{Model}

The $\beta$ delayed deuteron emission probability per time and
energy units, $dW/dE$, can be expressed \cite{N16,Baye} as
\begin{equation}
{{dW}\over{dE}}={{mc^2}\over{\pi^4v\hbar^4}}G^2_\beta
f(Q-E)B_{\rm GT}(E),
\end{equation}
where $m$ is the electron mass, $v$ is the relative velocity
between $\alpha$ and deuteron and $G_\beta=3.002\cdot 10^{-12}$
is the dimensionless $\beta$-decay constant. The phase space
factor, or Fermi integral, $f$ depends on the kinetic energy
$Q-E$ available for the electron and antineutrino. The mass
difference $Q$ between the initial and final particles is 2.03
MeV. The maximum total energy (including mass energy) is
$W_0=2.54$ MeV. The GT reduced transition probability reads
\begin{equation}
B_{\rm GT}(E)={{\lambda^2}\over{(2J+1)}}
\sum_{M M'\mu} \vert \langle\Psi_{J'M'}^{\alpha
d}\left\vert\right .
\sum_{j=1}^6t^j_{-}\sigma^j_{\mu}\left\vert\right .
\Psi_{JM}^{^6{\rm He}}\rangle \vert^2,
\label{BGT}
\end{equation}
where $\Psi^{\alpha d}$ and $\Psi^{^6{\rm He}}$ are the
microscopic wave functions of the $\alpha$ + d scattering state
and of the $^6$He ground state, respectively (we shall specify them
later), $\mbox{\boldmath $t$}^j$ and $\mbox{\boldmath $\sigma$}^j$ are
respectively the isospin operator and Pauli spin-matrices of
nucleon $j$, and $\lambda =1.26$ is the ratio of the
axial-vector to vector coupling constants. The half life $t_{1/2}$ for
the $\beta$-decay to the ground
state of $^6$Li is calculated with
\begin{equation}
(ft_{1/2})^{-1} = (2\ln 2) \pi^3  (mc^2/\hbar) G_{\beta}^2
B_{\rm GT} ({\rm g.s.}) ,\label{dWdE}
\end{equation}
where $B_{\rm GT} ({\rm g.s.})$ is given by
\begin{equation}
B_{\rm GT} ({\rm g.s.})={{\lambda^2}\over{(2J+1)}}\sum_{M M'\mu}
\vert \langle\Psi_{J'M'}^{^6{\rm Li}}
\left\vert\right .
\sum_{j=1}^6t^j_{-}\sigma^j_{\mu}\left\vert\right .
\Psi_{JM}^{^6{\rm He}}\rangle \vert^2.
\label{BGT0}
\end{equation}

In \cite{Csoto} a microscopic dynamical
model has been developed for the description of
the ground state of $^6$He. In the
present work we use the wave function of that model
\widetext
\begin{eqnarray}
\Psi_{JM}^{^6{\rm He}}
&=&\sum_{S,l_1,l_2,L}\Psi_{S,(l_1l_2)L}^{\alpha
(nn)}+\sum_{S,l_1,l_2,L}\Psi_{S,(l_1l_2)L}^{n (\alpha n)}+
\Psi _{S,L}^{tt}
\nonumber \\
&=&\sum_{S,l_1,l_2,L}\sum_{i=0}^{N_a-1}{\cal A}\left \{\left [
\left [\Phi ^\alpha_i(\Phi ^n\Phi^n)
\right ]_S
\chi ^{\alpha (nn)}_{i[l_1l_2]L}(\mbox{\boldmath $\rho $}_{nn},
\mbox{\boldmath $\rho $}_{\alpha (nn)})
\right ]_{JM}
\right \}
\nonumber \\
&+&\sum_{S,l_1,l_2,L}\sum_{i=0}^{N_a-1}{\cal A}\left \{\left [
\left [\Phi ^n(\Phi ^\alpha_i\Phi^n)\right ]_S
\chi ^{n (\alpha n)}_{i[l_1l_2]L}(\mbox{\boldmath $\rho $}_{\alpha n},
\mbox{\boldmath $\rho $}_{n(\alpha
n)})
\right ]_{JM}
\right \}\nonumber \\
&+&\sum_{S,L} {\cal A}\left \{\left [ [\Phi ^t\Phi ^t
]\raise-0.66ex\hbox{\scriptsize $S$}
\chi ^{tt}_L(\mbox{\boldmath $\rho $}_{tt})
\right ]_{JM} \right \}.
\label{wfn}
\end{eqnarray}
\narrowtext
\noindent
Here ${\cal A}$ is the intercluster antisymmetrizer, the
\mbox{\boldmath $\rho $} vectors are the different intercluster
Jacobi coordinates, and [\ ] denotes angular momentum coupling.
While $\Phi ^n$ is a neutron spin-isospin eigenstate, $\Phi^t$
is the antisymmetrized triton internal state in the harmonic
oscillator shell model with a single oscillator parameter.
The antisymmetrized ground state ($i=0$) and monopole excited
states ($i>0$) of the $\alpha$ particle are represented by the
wave functions
\begin{equation}
\Phi ^\alpha _i=
\sum_{j=1}^{N_a}A_{ij}\phi ^\alpha _{\beta _j},\ \
i=0,1,...,(N_a-1), \label{alpha}
\end{equation}
where $\phi ^\alpha _{\beta _j}$ is a translation invariant
shell--model wave function of the $\alpha$ particle with size
parameter $\beta _j$ and the $A_{ij}$ parameters are to be
determined by minimizing the energy of
the $\alpha$ particle \cite{Tang}. Similarly the size
parameter of the tritons is determined from the energy stabilization.
We choose the same parameters for
the wave function as in Ref. \cite{Csoto} with $N_a=3$. The following
$SL$ terms are included in the $S,(l_1l_2)L$ coupling scheme for the
$J^\pi=0^+$ ground state of $^6$He:
$\{\alpha (nn);0(00)0\}$, $\{\alpha (nn);1(11)1\}$,
$\{n(\alpha n);0(00)0\}$, $\{n(\alpha n);0(11)0\}$,
$\{n(\alpha n);1(11)1\}$, and $\{tt,00\}$. Putting (\ref{wfn})
into the six-nucleon Schr\"odinger equation, we
arrive at an equation for the intercluster relative motion functions
$\chi$. These functions are expanded in terms of products of
tempered Gaussian
functions $\exp (-\gamma_i \rho^2)$ \cite{Kamimura} with different
ranges $\gamma_i$ for each type of relative coordinate.
The expansion coefficients are determined
from a variational principle.

The Minnesota force \cite{Minnesota} with an exchange parameter
$u=0.92$ and a slightly modified spin-orbit component (see Ref.
\cite{Csoto}) reproduces very
well the experimental phase shifts of all the N + N and $\alpha$
+ N scattering states which take place in (\ref{wfn})
\cite{Csoto}. After this force choice there remains no free
parameter in
the model. This model gives for $^6$He an $\alpha$ + n + n
three-cluster separation
energy of 0.961 MeV, i.e. practically the experimental value
0.975 MeV. It was pointed out in Ref. \cite{Csoto} that
this value could not be reproduced without the
t + t component.
As the GT matrix element is sensitive
to the details of the $^6$He wave function up to 15-20 fm in
the $\rho_{\alpha (nn)}$ coordinate \cite{Baye}, our wave
function  should be correct in this region. To achieve this, we
enlarge the number of basis states which describe large
$\rho_{\alpha (nn)}$ separations in such a way that the range of
our last basis functions is $\sim 20$ fm in these relative motions.

To be consistent, we must employ the same model
parameters in the description of the $\alpha$ + d
bound and scattering states as for $^6$He. For example
the use of a different nucleon-nucleon interaction
for $\alpha$ + d would affect
the product of the spatial parts of the $^6$He and $\alpha$ + d
wave functions and modify the balance between the different parts
in the GT matrix element, and could lead to inaccurate results.
The Minnesota force reproduces the N + N effective range
parameters in the triplet even partial wave by assuming a pure
$^3$$S_1$ state, i.e. without tensor coupling with a $D$ state.
However, if a tensor force were taken into account
in addition to the Minnesota force,
it would play a very large role in this partial wave
in a coupled $^3$$S_1$--$^3$$D_1$ description \cite{Friar}.
This means that if a pure $^3$$S_1$ state reproduces the
experiment, then the effective interaction in this partial wave
is too strong. As was pointed out in \cite{Csoto}, the Minnesota
interaction  overbinds the ground state of $^6$Li by 1.2 MeV
when all possible angular momentum configurations are taken into
account. To avoid this, we keep only the $L=0$, $S=1$
configuration. The error we make with this truncation is
acceptable because only a $\sim 5\%$ $L\neq 0$ component
\cite{CsotoLovas} in $^6$Li is neglected, but in compensation
the asymptotic form of the wave funtion is more realistic.
The contribution to $B_{\rm GT}$ from the
$^6$He$\;(L=1)$$\rightarrow$$^6$Li transition would be
negligible ($\sim 0.1\%$) since the GT operator only connects
states with the same $L$. However, because of the lack of the
$L=1$ component, the weight
of the $L=0$ component is increased so that $B_{\rm GT}$ might be
overestimated by 5\% although the variational principle might partly
compensate this effect. Also the presence of $L\neq 0$
components in the $^6$Li wave function might
slightly shift the positions of its nodes. The importance of these errors
can be estimated from the calculation of $B_{\rm GT} ({\rm g.s.})$.

After these choices, our $^6$Li wave functions read with $L=0$,
$S=1$, and $J^{\pi}=1^+$,
\begin{eqnarray}
\Psi_{JM}^{^6{\rm Li}}
&=&\sum_{i=0}^{N_d-1}{\cal A}\left \{\left [ \left [\Phi
^\alpha \Phi ^d_i\right]\raise-0.66ex\hbox{\scriptsize $S$}\
\chi^{\alpha d}_{iL}(\mbox{\boldmath $\rho $}_{\alpha d})
\right ]_{JM}
\right \}\nonumber \\
&+&{\cal A}\left \{\left [ [\Phi ^t\Phi ^h
]\raise-0.66ex\hbox{\scriptsize $S$} \
\chi^{th}_L(\mbox{\boldmath $\rho $}_{th})
\right ]_{JM} \right \}
\label{li6}
\end{eqnarray}
for the ground state, and
\begin{eqnarray}
\Psi_{JM}^{\alpha d}
&=&{\cal A}\left \{\left [ \left [\Phi
^\alpha \Phi ^d_0\right]\raise-0.66ex\hbox{\scriptsize $S$}
\ g^{\alpha d}_{L}(E,\mbox{\boldmath $\rho $}_{\alpha d})
\right ]_{JM}\right \}\nonumber \\
&+&
\sum_{i=1}^{N_d-1}{\cal A}\left \{\left [ \left [\Phi
^\alpha\Phi ^d_i\right ]\raise-0.66ex\hbox{\scriptsize $S$}
\ \chi^{\alpha d}_{iL}(E,\mbox{\boldmath $\rho $}_{\alpha d})
\right ]_{JM}
\right \}\nonumber \\
&+&{\cal A}\left \{\left [ [\Phi ^t\Phi ^h
]\raise-0.66ex\hbox{\scriptsize $S$}
\ \chi^{th}_L(E,\mbox{\boldmath $\rho $}_{th})
\right ]_{JM} \right \}
\label{ad}
\end{eqnarray}
for the scattering states (h = $^3$He), where $E$ is the $\alpha$
+ d relative motion energy in the center of mass frame.
The normalization of $g$ is chosen consistently with
(\ref{dWdE}) as
\begin{eqnarray}
g^{\alpha d}_{0}(E,\mbox{\boldmath $\rho$}_{\alpha d})
&\rightarrow & Y_{00} (\hat{\mbox{\boldmath $\rho $}}_{\alpha
d}) \rho_{\alpha d}^{-1} \nonumber \\
&\times& \Big (F_0(k\rho_{\alpha d})\cos \delta
+G_0(k\rho_{\alpha d})\sin \delta \Big )
\end{eqnarray}
if $\rho_{\alpha d}\rightarrow\infty$, where $k$ is the wave
number, $F_0$ and $G_0$ are Coulomb
functions, and $\delta$ is the $s$-wave phase shift, at energy $E$.

The deuteron cluster being very distortable we use 5
basis states ($N_d=5$) for its description. The ground
state $\Phi ^d_0$ with $-2.20$ MeV energy and 2.1 fm (point nucleon)
rms radius, and four pseudostates $\Phi ^d_i$ are included in
(\ref{li6}) and (\ref{ad}). Because the distortion
of the $\alpha$ particle is weak in scattering processes,
we describe it with a single
stabilized size parameter.
Using the same force as for $^6$He, the ground state of
$^6$Li described by (\ref{li6}) provides the
experimental $\alpha$ + d cluster separation energy of 1.47 MeV.
It is remarkable that we can reproduce both the $^6$Li and $^6$He
ground states with the same force in this model. Notice that
other configuration choices underestimate the binding energy.
(i) Without the t + $^3$He component, $^6$Li is underbound by
more than 0.5 MeV. (ii) This result is {\em not} improved by
$\alpha$ distortion. (iii) With $\alpha$ distortion {\em and}
the t + $^3$He component, $^6$Li is also underbound but by the
smaller value 0.24 MeV; the role of t + $^3$He is weakened
because of a higher threshold. For the purpose of tests
described below, it is interesting to have another wave function
reproducing the experimental $^6$Li energy without the t +
$^3$He component. This can be achieved by refitting the exchange
mixture parameter to the value $u = 0.97$.

The scattering states are calculated from a Kohn-Hulth\'en
variational method for the $S$-matrix, which uses square
integrable basis functions matched with the correct scattering
asymptotics \cite{Kamimura}. To make the calculations
numerically stable, the matching radius must be chosen in the
10-15 fm region. In order to calculate $B_{\rm GT}$ analytically
up to 25-30 fm, the scattering wave function $g$
coming from the variational method is also expanded
in terms of square integrable tempered Gaussian functions.
The squared deviation between $g$ and this expansion is
variationally minimized up to 40 fm and becomes
less than $10^{-8}$. The $L=0$ $\alpha$ + d phase
shifts in the $E=0.1-5.0$ MeV relative center of mass energy
region are compared with experiment \cite{Jenney}
in Fig.\ \ref{fig0}. They agree within 0.1 degree with those obtained
with the potential of Ref. \cite{Kukulinps}, which is fitted to
experiment. Putting the $\Psi^{^6{\rm He}}$, $\Psi^{^6{\rm
Li}}$, and $\Psi^{\alpha d}(E)$ wave functions, coming from the
variational calculations, into (\ref{BGT}) and (\ref{BGT0}) we
can compute the desired $B_{\rm GT}$ and $B_{\rm GT}$(g.s.)
matrix elements. All calculations are performed analytically
with the aid of a symbolic computer language.

Comparing our model to the previous theoretical models
\cite{Descouvemont,Zhukov,Baye} we can say the following.
A common drawback of those three models is that the $\alpha$ + d
scattering states are obtained
from an $\alpha$ + d nucleus-nucleus
potential. Although this non-microscopic
potential model reproduces the correct $\alpha$ + d phase shifts,
the inner part of the scattering wave functions is not well
established. It may not be consistent with the $^6$He description.
In Ref. \cite{Descouvemont} $^6$He is represented by a pure $\alpha (nn)$
configuration. The dineutron component is therefore
overestimated in the $^6$He wave function, and hence in its halo
part. Consequently the external part of the GT matrix element
and $B_{\rm GT}$ are too large. In Ref. \cite{Zhukov}, the
results are rather insensitive to the use of an attractive or a
repulsive interaction in the $s$ wave, which both reproduce the
phase shifts. This insensitivity is rather surprising because
phase-equivalent potentials usually do not lead to the same
results for off-shell effects \cite{brem}.
As we can see e.g.\ in Fig. 3\ of Ref. \cite{Zhukov}, the repulsive
potential does not produce an internal node in the $\alpha$ + d
wave function, while the attractive potential produces such a node
(Fig.\ 2 of Ref. \cite{Descouvemont}). Therefore, the GT matrix
elements with the $^6$He wave function should be very different
from each other. This effect is probably hidden by the
orthogonalization procedure adopted by the authors of Ref.
\cite{Zhukov}. In Ref. \cite{Baye}, the $^6$He wave function is not
totally free for the variational method: in the halo, the spatial
part is restricted to small interneutron distances. As the
interaction, employed there, does not reproduce correctly the
n + n scattering and binds the dineutron, it might distort the halo
in a larger variational space. Moreover neglecting the $\sim
15\%$ $L\neq 0$ component in that wave function leads to an
enlargement of $B_{\rm GT}$. Finally, using a microscopic wave function
as initial state with an $\alpha$ + d potential wave function
as final state, even with the correct phase shifts, is not consistent.
The present model is essentially free of such defects. The description of
$^6$He is consistent with that of $^6$Li and of $\alpha$ + d,
the interaction satisfies conditions (i) to (vi) of Sec.\ I and
does not bind the dineutron, and $L\neq 0$ components are
explicitly included in the description of $^6$He.

\section{Results and discussion}

For $B_{\rm GT} ({\rm g.s.})$ we obtain 4.60$\lambda^2$ in our
model which corresponds to a half life $t_{1/2} \approx 835$ ms,
to be compared with the experimental value 806.7$\pm$1.5 ms
\cite{Ajzenberg}. In the test model, where the t + $^3$He
component is omitted in $^6$Li and the force is refitted, we get
4.48$\lambda^2$ for $B_{\rm GT} ({\rm g.s.})$ and 857 ms for
$t_{1/2}$. This example shows the importance of using the
same interaction in both the initial and final states of the
decay process: the error on the theoretical result
is doubled in the test calculation. The 3-4\% deviation
from experiment in the full model can be considered as a good
agreement. Of course the lack of $L \neq 1$ component in $^6$Li
probably reduces the difference between theory and experiment.
Nevertheless, as discussed in Ref. \cite{Lehman}, deviations as
large as 7-8\% could be accounted for by meson exchange current
corrections. Such corrections are typical of different weak
processes in few-nucleon systems (see e.g. Ref. \cite{Carlson}).

The calculated $dW/dE$ curves are shown in Fig.\ \ref{fig1}
together with the experimental points of Ref.\ \cite{Borge}. We
can see that both the full, consistent, model (solid curve) and
the test model (long-dashed curve) undershoot the
experimental points, but the results of the consistent model are
much closer to them. As for $B_{\rm GT}$(g.s), the
deviation from experiment is roughly twice as large in
the test model as in the full one. The error on $B_{\rm GT}$(g.s)
is typical of a process occurring at the normal nuclear scale,
i.e. at usual distances between nucleons in a nucleus. This
factor of two is still observed in $dW/dE$. However the 3-4\%
error on $B_{\rm GT}$(g.s.) in our best model is amplified to
60-70\% on $dW/dE$. This emphasizes again that the small $B_{\rm
GT}$ value in the $\beta$ decay to $\alpha$ + d is the
result of a balanced cancellation between two larger values with
different signs. The total branching ratio in the consistent
model is $6.0 \cdot 10^{-6}$. The branching ratio for deuterons
whose energy is above the experimental cutoff \cite{Borge} is
$3.1 \cdot 10^{-6}$. Let us note that the integration of the
experimental data in Fig.\ \ref{fig1} provides $5.4 \cdot
10^{-6}$, i.e. less than the value $(7.6 \pm 0.6) \cdot 10^{-6}$
quoted in Ref. \cite{Borge}.

To seek further after possible origins of errors, we changed the
number of basis functions between 8 and 25 in the critical
region of $\alpha (nn)$ relative motion in $^6$He, each basis
always ensuring the covering of this region up to 20-25 fm. Our
results remain stable within one percent. Next we checked the
asymptotics of our $^6$He wave function. In Fig.\ \ref{fig2} we
show the radial part of the $\chi ^{\alpha (nn)}_{0[00]0}$
relative motion function at fixed $\rho_{nn}=2$ fm. One
can see that the Gaussian cutoff starts only at roughly 30 fm.
The correct asymptotic behavior of the bound state wave function
of three neutral particles which have no bound binary
subchannels is given by \cite{Merkuriev}
\begin{equation}
\Psi \rightarrow \varrho^{-5/2}\exp (-\kappa\varrho)
\label{asymp}
\end{equation}
if $\varrho$$\rightarrow$$\infty$. Here $\varrho$ is the
hyperradius (defined by $\varrho^2 =\sum_iA_i r_i^2$,
where $\mbox{\boldmath $r$}_i$
is the center-of-mass coordinate of cluster $i$ with
respect to the center of mass of the three-cluster system)
and $\kappa
=(2m_N E_B/\hbar^2)^{1/2}$, where $E_B$ is
the three-cluster separation energy and $m_N$ the nucleon mass.
Since we are interested in the
$\alpha (nn)$ relative motion, $\varrho$ can be
expressed as $\varrho =\sqrt{{{4}\over{3}}\rho^2_{\alpha
(nn)}+{{1}\over{2}}\rho^2_{nn}}$. We checked that our $\chi
^{\alpha (nn)}_{0[00]0}$ relative motion
function satisfies the asymptotic form (\ref{asymp}) (the
other channels only slightly
contribute when $\rho_{nn}$ is small and
$\rho_{\alpha (nn)}$ is large).

Finally, to test the contribution coming
from the halo part of the $^6$He wave function,
we choose to expand the $\rho_{\alpha (nn)}$ space in (\ref{wfn})
with 10 basis functions which cover only the
0-10 fm region. Of course this truncation does not affect the
$^6$He energy at all. The corresponding asymptotic behavior of
$\chi ^{\alpha (nn)}_{0[00]0}$ is also shown in Fig.\
\ref{fig2}. The resulting $dW/dE$ is the short-dashed
curve in Fig.\ \ref{fig1}. As we can see, neglecting the
$8-10<\rho_{\alpha (nn)}<25$ fm part of the $^6$He wave function can
cause an order of magnitude error. This confirms that the
branching ratio is very sensitive to the halo part of the $^6$He
wave function, as was pointed out in Ref. \cite{Baye}.

The theoretical curve displays a satisfactory order of magnitude
although the data appear to be underestimated by about a factor
1.7 between 0.6 and 1.45 MeV. Its shape resembles the
experimental curve, but translated by about 200 keV toward lower
energies. As such a translation does not have any obvious
theoretical explanation, let us first discuss to which extent
experiment constraints the shape of the theoretical curve. The
data curve is significant only between 0.6 and 1.45 MeV. Below
this region, the experimental cutoff hides the energy dependence
of the results. Very few events (five) are observed above 1.45
MeV. Hence the real shape of $dW/dE$ and in particular the
location of its maximum remains open questions. As we now
discuss, these questions are of theoretical importance.

Most theoretical results (those of the present microscopic
calculation and of its two approximations displayed in Fig.\ \ref{fig1},
those of Refs. \cite{Descouvemont,Zhukov}, and several cases studied
in Ref. \cite{Baye}) display a common energy behavior with a maximum
between 0.3 and 0.4 fm. Only the magnitude of $dW/dE$ varies strongly.
Notable exceptions are given by the EH3 and EH results in Ref.
\cite{Baye} whose maximum is at or beyond 0.5 fm, and the EH2
result which displays a zero near 0.6 MeV in contradiction with
experiment. These non standard shapes follow from small
modifications in the external component of the GT matrix element
for a fixed internal component. In spite of their apparently
better agreement with the available data, the EH3 or EH shapes
cannot be considered as confirmed. Experimental data below
0.5 MeV (c.m.) would therefore provide information about the
nature of the cancellation mechanism between the internal
and external components of the GT matrix element \cite{Baye}.
They would discriminate between 'standard' curves with similar
shapes but varying magnitudes, and 'non standard' ones with a
maximum at larger energies which is compatible with -- but not
established by -- the presently available data. The standard
shape is well reproduced by the present microscopic calculation.
The 1.7 multiplicative factor might partly be due to the
difficulty of determining the absolute normalization of experiment.
The non-standard shapes require a very specific cancellation
mechanism and would indicate the need for further improvements
of the theory.

The discrepancy between theory and experiment for $dW/dE$
can possibly be attributed to the same mechanism
as the few-percent discrepancy in $B_{\rm GT}$(g.s.).
A rather small correction acting mainly on the internal
part of the matrix element which would provide a correct
$B_{\rm GT}$(g.s.) should modify much more strongly $dW/dE$
and might even affect its energy dependence. A good
candidate is provided by meson-exchange currents. Indeed
meson-exchange corrections might have the necessary order
of magnitude and should affect differently the
internal and external parts of the GT matrix elements since
the distances between the subsystems are different.
Rather small modifications of the internal part may result in 50
to 100\% modifications in $dW/dE$ as discussed in Ref.
\cite{Baye}, and as shown here by the test calculation. The fact
that the tensor force is not taken into account in the
description of the $\alpha$ + d scattering may also play some
role, but is probably less important than for the $^6$Li ground
state. Its introduction in the model would however
exclude the use of the Minnesota force and require the
difficult construction of a new force satisfying conditions
(i) to (vi) of Sec. I.

\section{Conclusion}

In summary, we have studied the $\beta$ delayed deut\-eron emission
from $^6$He in a microscopic cluster model. We have chosen the
Minnesota nucleon-nucleon interaction which reproduces the bulk
properties of the free clusters and gives very good
agreement with the experimental $N+N$ and $\alpha +N$ phase
shifts in all relevant partial waves. This interaction was shown
to provide a good overall description of the ground state of
$^6$He \cite{Csoto}. In the present work we point out that this
interaction also provides high quality results for the ground
state and $\alpha$ + d scattering states of $^6$Li. All these
results are obtained without any free parameter. The $^6$He,
$^6$Li, and $\alpha$ + d wave functions are used to calculate
the Gamow-Teller matrix element of the $\beta$ decay process.

In a calculation of the transition to the
$^6$Li ground state we emphasize the importance of using a consistent
description for the systems appearing in the initial and final
states, respectively. The results for the $\beta$ decay toward
continuum states show the strong sensitivity to the halo part of
the  $^6$He wave function predicted in Ref. \cite{Baye}.
Our best model gives a $\beta$ delayed deuteron spectrum close
to the experiment: the branching ratio is $3.1 \cdot 10^{-6}$ to
be compared with  $(7.6\pm 0.6)\cdot 10^{-6}$ \cite{Borge}, when
the experimental cutoff is assumed. The total theoretical
branching ratio is $6.0 \cdot 10^{-6}$. We argue that the
difference between the theoretical and experimental
$dW/dE$ has the same origin as the small difference between
the values for $B_{\rm GT} ({\rm g.s.})$ in spite of the fact
that it is much larger. We conjecture that corrections due to
meson-exchange currents are important enough to explain these
differences.

\acknowledgments

A.\ C.\ was supported by a Research Fellowship
from the Science Policy Office (Belgium) and by
OTKA grants Nos.\ 3010 and F4348 (Hungary). He is grateful
to the members of the Theoretical Nuclear Physics Department of the
Universit\'e Libre de Bruxelles for their kind hospitality, and
to M. Kruglanski for his help.

\raggedright

\begin{figure}
\caption{$\alpha$ + d phase shift for the $s$ wave obtained with the
microscopic wave function $\Psi_{JM}^{\alpha d}$ [Eq.
(\protect\ref{ad})]. The experimental points are taken from
\protect\cite{Jenney}. }
\label{fig0}
\end{figure}

\widetext
\begin{figure}
\caption{Transition probability $dW/dE$ per time and energy units
in the center of mass frame as a function of the center of mass
energy $E$. The experimental points are taken from Ref.\
\protect\cite{Borge}, while the statistical error bars are from Ref.\
\protect\cite{Baye}. The solid curve is the result of the present
calculation with the t + $^3$He component included in the
$\alpha$ + d wave function, while the long-dashed curve is the
result without this component. The short-dashed curve is obtained
by dropping the $\rho_{\alpha (nn)}> 10$ fm part of the $^6$He wave
function.}
\label{fig1}
\end{figure}

\narrowtext
\begin{figure}
\caption{Radial part of the $\chi ^{\alpha (nn)}_{0[00]0}$
relative motion function of the $^6$He wave
function (full line) calculated for $\rho_{nn}=2$ fm,
and its truncated approximation (short-dashed
curve) corresponding to the short-dashed curve in Fig.\
\protect\ref{fig1}.}
\label{fig2}
\end{figure}

\end{document}